\documentclass[aps,prd,notitlepage,showpacs,nofootinbib,superscriptaddress]{revtex4-1}

\pdfoutput=1
\usepackage[usenames,dvipsnames]{color}  
\usepackage{graphicx} 

\usepackage{bm}
\usepackage{bbold}
\usepackage{latexsym}
\usepackage{amsthm}
\usepackage{amsmath}
\usepackage{xfrac}
\usepackage{wrapfig}
\usepackage{amssymb}
\usepackage{cancel}
\usepackage{multirow}
\usepackage{pst-3d}                  
\usepackage{pst-grad}                
\usepackage{pst-node}                
\usepackage{pst-slpe}                
\usepackage{pst-plot}                %
\usepackage{pst-coil}                %
\usepackage{pst-math}                %
\usepackage{pstricks-add}            %
\usepackage{rotating}
\usepackage{geometry}
\usepackage{hhline}
\usepackage{amsmath}
\usepackage{float}
\usepackage{fancybox}
\usepackage[utf8]{inputenc}

\usepackage[colorlinks=true,citecolor=darkred,urlcolor=darkred, pdfborder={0 0 0}]{hyperref}
\usepackage[normalem]{ulem}

\usepackage{color,soul}
\setul{0.5ex}{0.3ex}
\setulcolor{blue}

\usepackage{braket}
\definecolor{linkcolor}{rgb}{0,0,0.5}
\allowdisplaybreaks
\allowdisplaybreaks[1]
\allowdisplaybreaks[2]

\def\vev#1{\left\langle #1\right\rangle}

\definecolor{greenLinks}{rgb}{0, 0.6, 0}
\definecolor{blueLinks}{rgb}{0, 0, 0.6}
\definecolor{redLinks}{rgb}{0.6, 0, 0}
\definecolor{tempText}{rgb}{0.55, 0.10,0.67}
\definecolor{eprintLinks}{rgb}{0.4, 0.4, 0.4}
\definecolor{journalLinks}{rgb}{0.6, 0, 0}
\newcommand {\ignore}[1]{}

\usepackage{soul}
\definecolor{mightnightblue}{RGB}{25,25,112}

\definecolor{brown}{rgb}{0.59, 0.29, 0.0}

\definecolor{darkred}{rgb}{0.6,0,0}

\def\SM{$\mathrm{ SU(3)_C \times SU(2)_L \times U(1)_Y }$ }

\def\vev#1{\left\langle #1\right\rangle}

\def\lsim{\mathrel{\rlap{\lower4pt\hbox{\hskip1pt$\sim$}}
    \raise1pt\hbox{$<$}}}
\def\gsim{\mathrel{\rlap{\lower4pt\hbox{\hskip1pt$\sim$}}
    \raise1pt\hbox{$>$}}}

\def\U1s{$\mathrm{U_{1}^{(a)}\otimes U_{1}^{(b)}\otimes U_{1}^{(c)}\otimes U_{1}^{(d)}\otimes U_{1}^{(e)}}$ }
\def\3211{$\mathrm{SU(3) \times SU(2)_L \times U(1)_R \times U(1)_{B-L}}$ }
\def\321{$\mathrm{SU(3) \times SU(2) \times U(1)}$ }
\def\422{$\mathrm{SU(4) \times SU(2) \times SU(2)_R}$ }

\newcommand{\AddrAHEP}{%
  AHEP Group, Institut de F\'{i}sica Corpuscular --
  C.S.I.C./Universitat de Val\`{e}ncia, Parc Cient\'ific de Paterna.\\
 C/ Catedr\'atico Jos\'e Beltr\'an, 2 E-46980 Paterna (Valencia), Spain}
 
 \newcommand{\AddrMPI}{Max-Planck-Institut f\"ur Kernphysik, Saupfercheckweg 1, 69117 Heidelberg, Germany\vspace{0.2cm}}
 
  \newcommand{\AddrWurzburg}{Institut f\"ur Theoretische Physik und Astrophysik, University of W\"urzburg, \\  Campus Hubland Nord, D-97074 W\"urzburg, Germany\vspace{0.2cm}}

\begin{document}

\title{\boldmath \color{BrickRed} Absolute neutrino mass scale and dark matter stability from flavour symmetry}

\author{Salvador Centelles Chuli\'{a}}\email{chulia@mpi-hd.mpg.de}
\affiliation{\AddrMPI}
\author{Ricardo Cepedello}\email{ricardo.cepedello@physik.uni-wuerzburg.de}
\affiliation{\AddrWurzburg}
\author{Omar Medina}\email{omar.medina@ific.uv.es}
\affiliation{\AddrAHEP}

\begin{abstract}
\noindent
We explore a simple but extremely predictive extension of the scotogenic model. We promote the \textit{scotogenic symmetry} $\mathbb{Z}_2$ to the flavour non-Abelian symmetry $\Sigma(81)$, which can also automatically protect dark matter stability. In addition, $\Sigma(81)$ leads to striking predictions in the lepton sector: only Inverted Ordering is realised, the absolute neutrino mass scale is predicted to be $m_\text{lightest} \approx 7.5 \times 10^{-4}$ eV and the Majorana phases are correlated in such a way that $|m_{ee}| \approx 0.018$ eV. The model also leads to a strong correlation between the solar mixing angle $\theta_{12}$ and $\delta_{CP}$, which may be falsified by the next generation of neutrino oscillation experiments. The setup is minimal in the sense that no additional symmetries or flavons are required.
\end{abstract}

\maketitle

\section{Introduction}

Motivated by two fundamental problems of particle and astroparticle physics, namely the origin of neutrino masses \cite{Minkowski:1977sc,Yanagida:1979as,GellMann:1980vs,mohapatra:1980ia,Schechter:1980gr,Schechter:1982cv, ma:1998dn,CentellesChulia:2018gwr, CentellesChulia:2020dfh} and the nature of dark matter \cite{Aghanim:2018eyx}, there's been a great effort to relate them within a single, predictive framework. Unarguably, they both point towards the presence of physics beyond the Standard Model (SM), presumably with the addition of new particles and symmetries that account for a mass mechanism for neutrinos, a viable dark matter candidate and its stability.

An \emph{economical} approach to combine all these appealing properties is to consider radiative neutrino mass models \cite{ Bonnet:2012kz, Farzan:2012ev, Sierra:2014rxa, Cepedello:2017eqf, Yao:2017vtm, Cepedello:2018rfh, Klein:2019iws, CentellesChulia:2019xky} (for a review see \cite{Cai:2017jrq}). In this kind of models, fields running in a loop generate neutrino masses, giving rise to two clearly distinguishable particle sectors, one of which can be regarded as a dark sector by means of a symmetry. The stability of the dark matter candidate, i.e. the lightest of the particles belonging to the dark sector, is determined by the transformation properties of the SM fields and the dark sector under symmetries \cite{Bonilla:2018ynb,CentellesChulia:2019gic,Srivastava:2019xhh}. In the most simple scenarios, the SM fields transform only under an invariant subgroup of the symmetry, while any particle beyond those of the SM not belonging to this subgroup will not be able to decay solely to the SM, i.e. it will be part of the dark sector. A popular implementation of this principle is the scotogenic model \cite{Ma:2006km} and its many variants (see for instance \cite{Rojas:2018wym,Kang:2019sab,Leite:2019grf, Barreiros:2020gxu,Borah:2021rbx,Escribano:2021ymx,Mandal:2021yph,Sarma:2022bhl,Ma:2022bfa,Sarazin:2021nwo}).

While a large number of models built following the described approach are consistent with experimental data from neutrino oscillations and bounds from dark matter searches, there are further unknowns about fundamental particles that are also important to address. The SM lacks a suitable theoretical explanation for the masses and the mixing pattern of fermions. Furthermore, the majority of input parameters of the SM are directly related to this \emph{flavour puzzle}. The lepton mixing angles, being large and with a completely different structure in comparison to their analogues in the quark sector, manifest the lack of a first principle explanation of the flavour phenomenology \cite{Kajita:2016cak, McDonald:2016ixn, ParticleDataGroup:2020ssz}. Here is where \emph{flavour symmetries} can play a major role in explaining such mixing patterns and mass hierarchies.\footnote{While there is a vast bibliography on this topic, we direct the interested readers to the reviews \cite{King:2015aea,Feruglio:2019ybq}.} By means of imposing a flavour symmetry between the three generations it's possible to predict strong correlations between different observables. This is essential for a flavour symmetry model to be verifiable.

In this paper we build a model for radiative neutrino masses with a flavour symmetry $\Sigma(81)$. We focus on such a discrete group due to an interesting feature: $\Sigma(81)$ contains a non-trivial subgroup formed by the singlets and one of the triplet representations. This ensures, as we will show in section \ref{sec:dm}, that for a reasonable choice of the transformation properties of the field content under the flavour symmetry, one can straightforwardly obtain a stable dark matter candidate. Thus, providing a \emph{natural} framework to account for dark matter stability along with light radiative Majorana neutrino masses through a scotogenic-like mechanism. Other works with \emph{flavoured stability} are, for example, \cite{Hirsch:2010ru,Lavoura:2012cv,Boucenna:2012qb,Ma:2019iwj,deAnda:2021jzc}. Other works with a flavour group $\Sigma(81)$ are for example \cite{BenTov:2012xp,Hagedorn:2008bc}.

A more conventional role played by $\Sigma(81)$ symmetry is to strongly constrain the structure of the mass matrices of fermions, leading to strong predictions that can be tested in the following years by the next generation of neutrino oscillation \cite{abe2014long,Hyper-Kamiokande:2016srs,DUNE:2020jqi,JUNO:2015zny,Cao:2014bea}, and neutrinoless double beta decay experiments \cite{KamLAND-Zen:2016pfg, GERDA:2018zzh, Agostini:2017iyd, Alduino:2017ehq, Arnold:2016qyg, Albert:2014awa}. While the idea of imposing a flavour symmetry is certainly not new, we will show that our setup has a series of attractive and unique features, namely explaining the lepton mixing pattern, as well as predicting the absolute mass scale of neutrinos, their ordering and the Majorana phases, and therefore leading to a definite prediction for neutrinoless double beta decay ($0\nu ee$). Moreover, this is obtained without the need of extra \textit{flavons}, i.e. extra scalars that further break the flavour symmetry. In our setup the breaking of the flavour symmetry is done by extending the number of Higgs doublets, as a variant of a 3HDM and giving them non-trivial charges under $\Sigma(81)$. \\

The paper is structured as follows: in section~\ref{sec:model} we present the model setup, i.e. the field content, the charges under the SM gauge group and flavour symmetry and discuss some of its most important attributes. In section~\ref{sec:pred} we delve into its most important phenomenological predictions: absolute neutrino mass scale and ordering, strong correlations between oscillation observables and the implications for neutrinoless double beta decay. In section~\ref{sec:dm} we explicitly flesh out the non-Abelian stability mechanism provided by $\Sigma(81)$. The paper then closes with a short summary and conclusions. Details about the symmetry group $\Sigma(81)$ are relegated to Appendix~\ref{sec:Sigma81}. 

\section{The model setup}
\label{sec:model}

We extend the Standard Model gauge symmetry \SM by a global, discrete flavour group $\Sigma(81)$. This group is of the type $\Sigma(3N^3)$ and contains 9 singlets and 4 complex triplets, denoted as $\mathbf{1}_{(i,j)}$ with $i,j=0,1,2$ and $\mathbf{3}_X (\mathbf{\bar{3}}_X)$ with $X = A,B,C,D$ (see Appendix \ref{sec:Sigma81} for details). The irreducible representations $\mathbf{3}_D$ ($\mathbf{\bar{3}}_D$), together with the singlets, form a closed set under tensor products, implying that if every Standard Model field transforms as $\mathbf{3}_D$, $\mathbf{\bar{3}}_D$ or as one of the singlets, then any field transforming as $\mathbf{3}_{A,B,C}$ and their conjugates, will belong to the \textit{dark sector}. The lightest among them will then be a dark matter candidate. This relation between $\Sigma(81)$ and dark matter will be further discussed in section~\ref{sec:dm}.

\begin{table}[t]
\begin{center}
\begin{tabular}{| c || c | c | c |}
  \hline
&   \hspace{0.1cm}  Fields     \hspace{0.1cm}       &  \hspace{0.4cm}  \SM     \hspace{0.4cm}    & \hspace{0.4cm}   $\Sigma(81)$            \hspace{0.4cm}                              \\
\hline \hline
\multirow{3}{*}{ \begin{turn}{90} Visible \end{turn} } 
&   $L$        	         &   ($\mathbf{1}, \mathbf{2}, {-1/2}$)     &   $\mathbf{3}_D$        \\
&   $\ell_{R}$           &   ($\mathbf{1}, \mathbf{1}, {-1}$)       &   $\mathbf{\bar{3}}_D$  \\
&   $H$   	             &   ($\mathbf{1}, \mathbf{2}, {1/2}$)      &   $\mathbf{\bar{3}}_D$  \\
\hline \hline                                                                             
\multirow{3}{*}{ \begin{turn}{90} Dark \end{turn} }                          
& $N_{L, R}$  	      	 &  ($\mathbf{1}, \mathbf{1}, {0}$)         &  $\mathbf{3}_A$         \\
& $\eta$  	       	 &  ($\mathbf{1}, \mathbf{2}, {1/2}$)       &  $\mathbf{3}_A$         \\
& $\phi$  	       	 &  ($\mathbf{1}, \mathbf{2}, {1/2}$)       &  $\mathbf{\bar{3}}_A$   \\
    \hline
  \end{tabular}
\end{center}
\caption{\footnotesize Particle content and symmetry transformation properties under the SM gauge group and the flavour symmetry $\Sigma(81)$. Note that the fields of the visible sector transform as $\mathbf{3}_D$, $\mathbf{\bar{3}}_D$ or  $\mathbf{1}_{(i, j)}$, while the dark sector transforms as $\mathbf{3}_A$ or $\mathbf{\bar{3}}_A$. The lightest particle of the dark sector will be automatically stable. See text for details.}
 \label{tab:charges}
\end{table}

The field content of the SM is extended by adding a vector-like singlet $N$ and two Higgs-like scalars, transforming non-trivially under $\Sigma(81)$. All the fields and charges are given in table~\ref{tab:charges}. Comparing to the original scotogenic model \cite{Ma:2006km}, new fields were also required to generate neutrino masses at one-loop with $\Sigma(81)$. While for the simple $\mathbb{Z}_2$ symmetry of the scotogenic model, any product of an odd field under $\mathbb{Z}_2$  times itself transforms as a singlet under $\mathbb{Z}_2$, this is not the case for any of the triplet representations of $\Sigma(81)$. For this reason, one needs to promote the right-handed neutrino to a vector-like fermion and, on a similar footing, two copies of the inert doublet Higgs are required, $\eta$ and $\phi$. For simplicity, we split the most relevant parts of the Lagrangian as,
\begin{equation} \label{eq:lagrangians} 
    \mathcal{L} \; \supset \;
        \mathcal{L}^{V}_\text{Y} + \mathcal{L}^{D}_\text{Y} - \mathcal{V}_\text{s} \, ,
\end{equation}
where the scalar potential is further divided into parts, for convenience, as $\mathcal{V}_s = \mathcal{V}_\nu + \mathcal{V}_\text{soft} + ...$ . The first part of the potential contains the scalar interactions that enter in the neutrino mass, the second of soft breaking terms of mass dimension 2 and ``$...$'' denotes the rest of the usual four-scalar interactions, that are not interesting for the purpose of our discussion. The terms in $\mathcal{V}_\text{soft}$ are of the form,
\begin{equation}
     \mathcal{V}_\text{soft} = \mu_{ij}^2 \, H_i^\dagger H_j \, .
\end{equation}

These terms are necessary in order to satisfy phenomenological bounds. In the limit $\mu_{ij} \rightarrow 0$, which we will call the ``symmetric limit'', the allowed VEV alignments will be highly restricted by the $\Sigma(81)$ symmetry. A preliminary analysis of the scalar potential, solving the tadpole equations, always yields highly symmetrical VEV alignments in this limit, for example $(v_1, v_2, v_3) = v \, (1, 0, 0)$ or $v \, (1, 1, 1)$. However, from the mass matrices shown in sections~\ref{sec:CL} and \ref{sec:numass}, it's evident that realistic lepton mixing and masses cannot be realized from such symmetric alignments. Including the $\mu_{ij}$ terms will add nine new parameters to the tadpole equations, allowing enough deviations from the symmetric limit to get a realistic lepton mixing pattern. In particular, the alignment that has been obtained from the phenomenological analysis is a perturbation from the tadpole equation solution, 
\begin{equation} \label{eq:vev align}
v(1, 0, 0) \rightarrow v(1, \epsilon_1, \epsilon_2)\quad \text{with} \quad \epsilon_1, \epsilon_2 \sim \mathcal{O}(10^{-2}).
\end{equation}
It is worth mentioning that such alignment approximately preserves a residual $\mathbb{Z}_3$ symmetry, which originates due to the invariance of the tadpole equation solution $v(1, 0, 0)$ under the $a'$ generator of $\Sigma(81)$ in the $\mathbf{\bar{3}}_{D}$ representation, as can be seen directly from equation (\ref{eq:3DRep}) in Appendix \ref{sec:Sigma81}.
On the other hand, the symmetric limit faces other phenomenological challenges. Strong FCNCs are expected in this type of models, as well as deviations from the Standard Model fermion-Higgs interactions. Following \cite{Georgi:1978ri}, after we have chosen a particular VEV alignment, we can rotate to the Higgs basis $H = \sum_i \frac{v_i}{v} H_i$. In this basis, only one doublet $H$ has a non-zero VEV, while the other two orthogonal combinations remain VEV-less. Then, the diagonal $\mu$ terms for these two doublets can be taken to be arbitrarily large, thus effectively decoupling them from the rest of the model without affecting the VEV structure. The doublet $H$ will be SM-like. Finally, for a more detailed analysis of the 3HDM scalar potential, see for example \cite{deMedeirosVarzielas:2021zqs,Ivanov:2012ry,Keus:2013hya,Ivanov:2014doa,Maniatis:2014oza,Vergeest:2022mqm,Hernandez:2021iss, deMedeirosVarzielas:2022kbj}.

\subsection{Charged lepton masses}
\label{sec:CL}

In this section, we derive the mass matrix for the charged leptons given the particle content of table~\ref{tab:charges}. The relevant piece of the Lagrangian in equation \eqref{eq:lagrangians} is the first term, which contains the Yukawa interaction terms among fields of the visible sector. To make the derivation clearer, all terms have been expanded in $\Sigma(81)$ components, for example, $L = ( L_1, L_2, L_3 )^\text{T}$, and similarly for the other triplets, following the tensor products given in the second edition of the book ``An Introduction to Non-Abelian Discrete Symmetries for Particle Physicists'' \cite{Kobayashi:2022moq} (see also Appendix~\ref{sec:Sigma81} for more details). In this way, it's made explicit in the Lagrangian itself that several contractions may lead to a singlet under $\Sigma(81)$. For instance, three triplets $\mathbf{3}_D$ have three different contractions to an invariant singlet $\mathbf{1}_{(0,0)}$. 

The Yukawa interactions among the visible sector are given by,
\begin{eqnarray} \label{eq:LVY}
    \mathcal{L}^{V}_\text{Y} \; &=&  \; Y^{e}_1 \sum_{i=1}^{3} \bar{L}_i \ell_{R_i} H_i
    \\ &+& Y^{e}_2 \left( \bar{L}_1 \ell_{R_3} H_2 + \bar{L}_2 \ell_{R_1} H_3 + \bar{L}_3 \ell_{R_2} H_1 \right)
    \nonumber
    \\ &+& Y^{e}_3 \left( \bar{L}_1 \ell_{R_2} H_3 + \bar{L}_2 \ell_{R_3} H_1 + \bar{L}_3 \ell_{R_1} H_2 \right)
    \nonumber
    \\ \nonumber &+& \text{h.c.} \, ,
\end{eqnarray}
where $SU(2)_L$ indices and contractions have been omitted for simplicity. After electroweak symmetry breaking (EWSB) the Lagrangian $\mathcal{L}^{V}_\text{Y}$ in last equation gives rise to the mass matrix for charged leptons,
\begin{equation} \label{eq:Mcl}
    M_e = \frac{1}{\sqrt{2}} \, \left( 
        \begin{array}{ccc}
            Y_1^e \, v1 & Y_3^e \, v3 & Y_2^e \, v2 \\
            Y_2^e \, v3 & Y_1^e \, v2 & Y_3^e \, v1 \\
            Y_3^e \, v2 & Y_2^e \, v1 & Y_1^e \, v3
        \end{array}
    \right),
\end{equation}
in the basis $\{ (L_1,L_2,L_3), \, (\ell_{R_1}, \ell_{R_2}, \ell_{R_3}) \}$, with the vacuum expectation values of the Higgs defined as, $\vev{H_i} = v_i/\sqrt{2}$ and $\sum_i v_i^2 = v_\text{SM}^2$ the Standard Model VEV. The mass matrix $M_e$ is then diagonalised by the unitary rotations $U_\ell$ and $V_\ell$ as,
\begin{eqnarray}
\label{eq:cldiag}
    \hat{M}_e = U_\ell^\dagger M_e V_\ell, \quad \text{with} \quad L \rightarrow U_\ell \, L, \hspace{0.5cm} \ell_R \rightarrow V_\ell \, \ell_R \, , \quad  \hat{M}_e = \text{diag} (m_e, m_\mu, m_\tau) \, .
\end{eqnarray}

\subsection{Neutrino masses}
\label{sec:numass}

\begin{figure}[tb]
\centering
\includegraphics[width=0.6\textwidth]{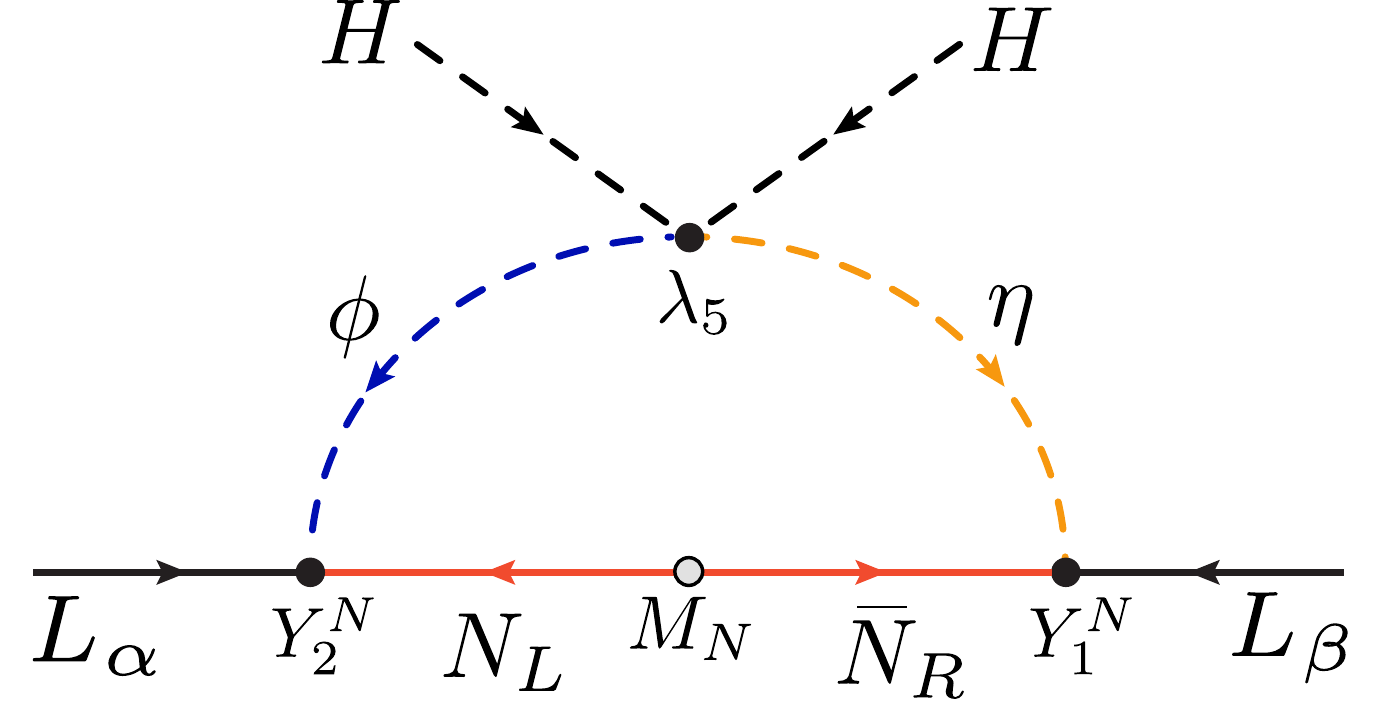}
\caption{Diagram generating neutrino masses at one-loop in our model, analogous to the original scotogenic model. Fields running in the loop are charged as $\mathbf{3}_A$ or $\mathbf{\bar{3}}_A$ under $\Sigma(81)$, while fields in the external legs (SM) transform as $\mathbf{3}_D$ or $\mathbf{\bar{3}}_D$.
  \label{fig:nudiagram}
  }
\end{figure}

The term $\mathcal{L}^{D}_\text{Y}$ in the RHS of equation \eqref{eq:lagrangians},  describes the interactions with the fields of the dark sector. This piece, together with the scalar potential, will give rise to the one-loop neutrino mass diagram depicted in figure~\ref{fig:nudiagram} and its corresponding mass matrix. This term in the Lagrangian is given by,
\begin{eqnarray} \label{eq:LDY}
    \mathcal{L}^{D}_\text{Y} \; &=&  \; M_N \left( \textcolor{violet}{\bar{N}_{L_1} N_{R_1}} + \textcolor{blue}{\bar{N}_{L_2} N_{R_2}} + \textcolor{OliveGreen}{\bar{N}_{L_3} N_{R_3}} \right)
   \\ &+& Y^{N}_1 \left( \textcolor{blue}{L_1 \bar{N}_{R_2} \eta_1}  +  \textcolor{OliveGreen}{L_2 \bar{N}_{R_3} \eta_2} + \textcolor{violet}{L_3 \bar{N}_{R_1} \eta_3} \right)
    \nonumber
    \\ &+& Y^{N}_2 \left( \textcolor{violet}{L_1 N_{L_1} \phi_2} + \textcolor{blue}{L_2 N_{L_2} \phi_3} + \textcolor{OliveGreen}{L_3 N_{L_3} \phi_1} \right)
    \nonumber
    \\ \nonumber &+& \text{h.c.} \, .
\end{eqnarray}
The relevant scalar couplings, analogous to the $\lambda_5$ interaction from the original scotogenic model \cite{Ma:2006km}, are
\begin{eqnarray} \label{eq:Vnu}
    \mathcal{V}_\nu \; &\supset& \lambda_5^{(1)} \left[ \textcolor{OliveGreen}{(H_1 \eta_2^\dagger) (H_1 \phi_1^\dagger)} + \textcolor{violet}{(H_2 \eta_3^\dagger) (H_2 \phi_2^\dagger)} + \textcolor{blue}{(H_3 \eta_1^\dagger) (H_3 \phi_3^\dagger)} \right]
    \nonumber
    \\ &+& \lambda_5^{(2)} \left[ \textcolor{blue}{(H_1 \eta_1^\dagger) (H_2 \phi_3^\dagger)} + \textcolor{violet}{(H_1 \eta_3^\dagger) (H_3 \phi_2^\dagger)} + \textcolor{OliveGreen}{(H_2 \eta_2^\dagger) (H_3 \phi_1^\dagger)} \right]
    \nonumber
    \\ &+& \text{h.c.} \, .
\end{eqnarray}
The expansion in components of $\Sigma(81)$ makes explicit that not every entry of the neutrino mass matrix will be generated. In fact, there are only six possible diagrams with different $\Sigma(81)$ components outside the loop and running in it. After EWSB, the resultant neutrino mass matrix is of the form,
\begin{equation} \label{eq:mnu}
    M_\nu \sim \frac{1}{2} \, \left( 
        \begin{array}{ccc}
             0  &  \textcolor{blue}{C_1 v_3^2 + C_2 v_1 v_2}  &  \textcolor{violet}{C_1 v_2^2 + C_2 v_1 v_3} \\
            \textcolor{blue}{C_1 v_3^2 + C_2 v_1 v_2}  &  0  &  \textcolor{OliveGreen}{C_1 v_1^2 + C_2 v_2 v_3} \\
            \textcolor{violet}{C_1 v_2^2 + C_2 v_1 v_3}  &  \textcolor{OliveGreen}{C_1 v_1^2 + C_2 v_2 v_3}  &   0
        \end{array}
    \right) \, .
\end{equation}
For the sake of clarity, we have assigned colours to each entry of the matrix and to its corresponding terms in the Lagrangian $\mathcal{L}^{D}_\text{Y}$ and the scalar potential $\mathcal{V}_\nu$ in equations \eqref{eq:LDY} and \eqref{eq:Vnu} respectively. The coefficients $C_a$ are obtained by computing the different diagrams of the type of figure~\ref{fig:nudiagram} that contribute,
\begin{eqnarray} \label{eq:Ccoefs}
    C_a \sim \frac{1}{16\pi^2} \, \frac{ \lambda_5^{(a)} \, (Y^N_1) \, (Y^N_2)}{ M_N } \, .
\end{eqnarray}
A very remarkable feature of the UV-realisation with $\Sigma(81)$ that we present here, is the fact that the neutrino matrix is exactly traceless with vanishing diagonal entries. This feature is protected by the symmetry and yields several strong predictions in the neutrino sector, as we will discuss in the next section. The matrix in equation \eqref{eq:mnu} coefficients $C_a$ correspond to the dominant contribution. The neutrino mass matrix is, in general, given by,
\begin{eqnarray} \label{eq:Mnucomp}
    (M_\nu)_{\alpha\beta} = \frac{1}{16\pi^2} \, (Y^N_1)_{\beta i j} \, (Y^N_2)_{\alpha i j} \, M_{N} \sum_{X=R,I} \sigma_X \, (U_X^{\alpha})_{1i} \, (U_X^{\alpha})_{i2} \, B_0( 0, M_N, m_{X_i}^2 ) \, ,
\end{eqnarray}
 where $\sigma_{R,I} = \pm 1$. The expression for the neutrino mass matrix \eqref{eq:Mnucomp} is very similar to that of the original scotogenic model, where after electroweak symmetry breaking the neutral part of the scalar doublet in the loop splits into its $CP$-even and $CP$-odd components (denoted as $R$ and $I$, respectively) due to the quartic coupling $\lambda_5$. The result is the sum of two $B_0$ Passarino-Veltman loop functions \cite{Passarino:1978jh} with a relative minus sign. Also, similar to the generalised scotogenic models with several scalars \cite{Escribano:2020iqq}, the mixing among the different scalar doublets in the loop need to be considered. The main subtlety is that, given the flavour symmetry $\Sigma(81)$, not every coupling is allowed. The only non-zero Yukawa couplings are $(Y_1^N)_{112} = (Y_1^N)_{223} = (Y_1^N)_{331} = Y_1^N$ and $(Y_2^N)_{121} = (Y_2^N)_{232} = (Y_2^N)_{313} = Y_2^N$. While the mass matrices mixing the neutral components of the scalars can be trivially obtained from \eqref{eq:Vnu}, with diagonalising matrices $U^\alpha_R$ and $U^\alpha_I$, for the $CP$-even and odd components respectively, in the basis $( \eta_\alpha, \phi_k )$. Note that again $\Sigma(81)$ only allows the mixing among specific pairs of $\eta$ and $\phi$ (see the scalar potential \eqref{eq:Vnu}).

 It is worth noting that while lepton flavour is violated in the neutrino sector, the usual dominant one-loop contribution to cLFV, mediated directly by $Y_i^N$, is absent. The Yukawa structure \eqref{eq:LDY} leads to a diagonal contribution proportional to $(|Y_1^N|^2 + |Y_2^N|^2)$ and the charged lepton masses. Consequently, any cLFV process, like $\mu \rightarrow e \gamma$, is suppressed.

\section{Predictions} \label{sec:pred}

The neutrino mass matrix in equation (\ref{eq:mnu}) is diagonalised as,
\begin{eqnarray}
\label{eq:nudiag}
     U_\nu^T \, M_\nu  \, U_\nu \, = \, \text{diag}(m_1, m_2, m_3) \, ,
\end{eqnarray}
where $U_\nu$ is the neutrino unitary mixing matrix and $m_i$ are the neutrino masses. In the Normal Ordering case $m_3 > m_2 > m_1$, while in the Inverted Ordering case $m_2 > m_1 > m_3$. Considering both equations \eqref{eq:cldiag} and \eqref{eq:nudiag} we obtain the lepton mixing matrix,
\begin{equation}
    U_\text{lep} = U_\ell^\dagger\, U_\nu \, .
\end{equation}
$U_\text{lep}$ is constrained by neutrino oscillation experiments. We choose the so-called symmetric parametrisation of a general unitary matrix \cite{Schechter:1980gr, Rodejohann:2011vc},
\begin{equation}
U_\text{lep} = P(\delta_1, \delta_2, \delta_3) \, U_{23}(\theta_{23}, \phi_{23}) \, U_{13} (\theta_{13}, \phi_{13}) \, U_{12}(\theta_{12}, \phi_{12}) \, ,
\end{equation}
where $P(\delta_1, \delta_2, \delta_3)$ is a diagonal matrix of unphysical phases and the $U_{ij}$ are complex rotations in the $ij$ plane, as for example,
\begin{equation}
    U_{23} (\theta_{23}, \phi_{23}) = \left(\begin{matrix}
        1 & 0 & 0 \\
        0 & \cos\theta_{23} & \sin\theta_{23}\, e^{-i \phi_{23}} \\
        0 & -\sin\theta_{23} \,e^{i \phi_{23}} & \cos\theta_{23}
        \end{matrix} \right) \,.
\end{equation}
The phases $\phi_{12}$ and $\phi_{13}$ are relevant for neutrinoless double beta decay while the combination $\delta_{CP} = \phi_{13} - \phi_{12} - \phi_{23}$ is the usual Dirac $CP$ phase measured in neutrino oscillations.

Before going into the numerical results, let us note an interesting analytical property of the matrix \eqref{eq:mnu}. The shape of this mass matrix, due to the $\Sigma(81)$ flavour symmetry, implies that the neutrino masses satisfy the relation,
\begin{equation}
\label{eq:sumrule}
\frac{1}{2} \, \sum \, m_i \, = \, m_\text{heaviest} \, ,
\end{equation}
where $m_\text{heaviest} $ is the heaviest neutrino mass. Equation~\eqref{eq:sumrule} is actually a general prediction for a complex, symmetric, diagonal-less neutrino mass matrix. If we call such a mass matrix $A$ and define it in general as,
\begin{equation}
\label{eq:matA}
A = \left(\begin{matrix}
0 & a & b \\
a & 0 & c \\
b & c & 0
\end{matrix}\right), \hspace{0.5cm} \text{with } a, \, b, \, c \in \mathbb{C} \, ,
\end{equation}
diagonalised as usual by
\begin{eqnarray}
U^T A \,U &=& m_d = \text{diagonal}(m_1, m_2, m_3) \, , \\
U^\dagger A^\dagger A \, U &=& m_d^2 \, ,
\end{eqnarray}
where $m_{d}$ is real, diagonal and positive. With this definition the traces of the matrices $A^\dagger A$ and $(A^\dagger A)^2$ can be computed straightforwardly,
\begin{eqnarray}
Tr(A^\dagger A) = 2 (|a|^2+|b|^2+|c|^2) = m_1^2+m_2^2+m_3^2 \, , \\
Tr\left[(A^\dagger A)^2\right] = 2 (|a|^2+|b|^2+|c|^2)^2 = m_1^4+m_2^4+m_3^4 \, .
\end{eqnarray}
These traces fulfill the general relation,
\begin{equation}
    \frac{1}{2} \left[ Tr( A^\dagger A )\right]^2 = Tr \left[ (A^\dagger A)^2 \right] \, ,
\end{equation}
which translated to the mass eigenvalues reads,
\begin{equation}
    m_3^2 = (m_1 \pm m_2)^2 \, .
\end{equation}
Since $m_i$ are real and positive, only one solution survives after specifying the ordering. In particular,
\begin{eqnarray}
m_3^\text{NO} =m_1^\text{NO}+m_2^\text{NO} \, , \label{eq:sumruleNO}
\\
m_2^\text{IO} =m_1^\text{IO}+m_3^\text{IO}
\label{eq:sumruleIO}  \, ,
\end{eqnarray}
or in general, irrespective of the ordering, the sum rule \eqref{eq:sumrule}.

Neutrino oscillations measure the mass squared differences of neutrino masses \cite{IceCube:2019dqi,Super-Kamiokande:2017yvm,T2K:2021xwb,alex_himmel_2020_3959581,DayaBay:2018yms,KamLAND:2010fvi}, which in combination with the mass sum rules \eqref{eq:sumruleNO} and \eqref{eq:sumruleIO} lead to the prediction of the absolute scale of the neutrino masses:
\begin{eqnarray}
m_\text{lightest} ^\text{NO} \, &\approx& \,  2.8 \times 10^{-2} \, \text{eV} \, ,
\\
m_\text{lightest} ^\text{IO} \, &\approx& \, 7.5 \times 10^{-4} \, \text{eV} \, .
\end{eqnarray}
Both values are well below cosmological bounds \cite{Planck:2018vyg} and direct measurements of neutrino mass \cite{KATRIN:2021uub,KATRIN:2022ayy}. However, they may be probed with astrophysical sources \cite{Kyutoku:2017wnb}.

Note, however, that the neutrino mass matrix in equation \eqref{eq:mnu} is more restricted than the matrix in equation \eqref{eq:matA}. In particular, the strong hierarchy in the masses of the charged leptons implies a strong hierarchy between the VEVs of the Higgs doublets, further restricting the neutrino mass matrix. We have performed a numerical scan and found the following results and predictions for both orderings.

\subsection{Inverted ordering}

\begin{figure}[t]
    \centering \includegraphics[width=0.9\textwidth]{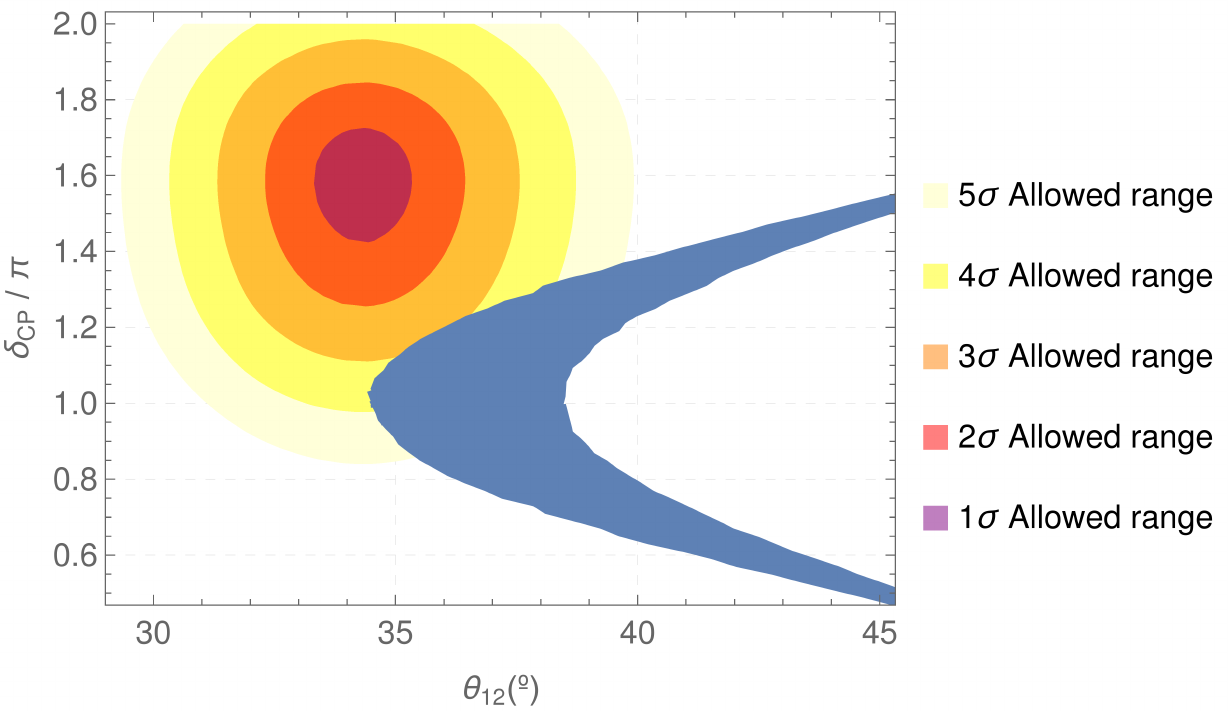}
    \caption{Correlation between $\theta_{12}$ and $\delta_{CP}$ in the IO case when the other observables are fitted inside their $3 \sigma$ experimental ranges: charged lepton masses, neutrino squared mass differences, $\theta_{13}$ and $\theta_{23}$. The $3\sigma$ tension in the combined $\theta_{12}-\delta_{CP}$ plane may be relieved if $\delta_{CP}$ is measured to be around $CP$ conserving values. In that case, $\theta_{12}$ would lie in its $1\sigma$ experimental region. $\chi^2$ profiles extracted from the global fit \cite{deSalas:2020pgw}.}
    \label{fig:IO12vdelta}
\end{figure}

In the Inverted Ordering case, a strong correlation appears between $\theta_{12}$ and $\delta_{CP}$ when the charged lepton masses, neutrino masses, $\theta_{13}$ and $\theta_{23}$ are fitted to their experimental values. The model can accommodate all oscillation observables inside their $3\sigma$ ranges with a slight tension in the $\theta_{12}$ vs $\delta_{CP}$ plane, see figure~\ref{fig:IO12vdelta}. However, the best fit value of $\delta_{CP}$ is very sensible to new data sets. An update from the Nova collaboration \cite{alex_himmel_2020_3959581} may change the picture in 2022. 

\begin{figure}[t]
    \centering
        \includegraphics[width=0.45\textwidth]{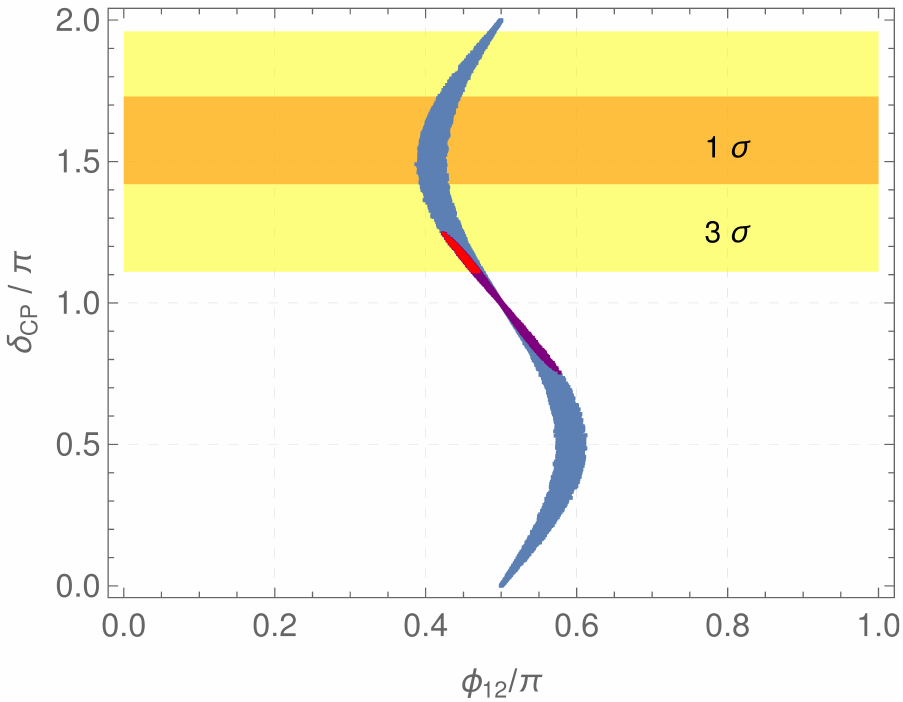}
        \includegraphics[width=0.45\textwidth]{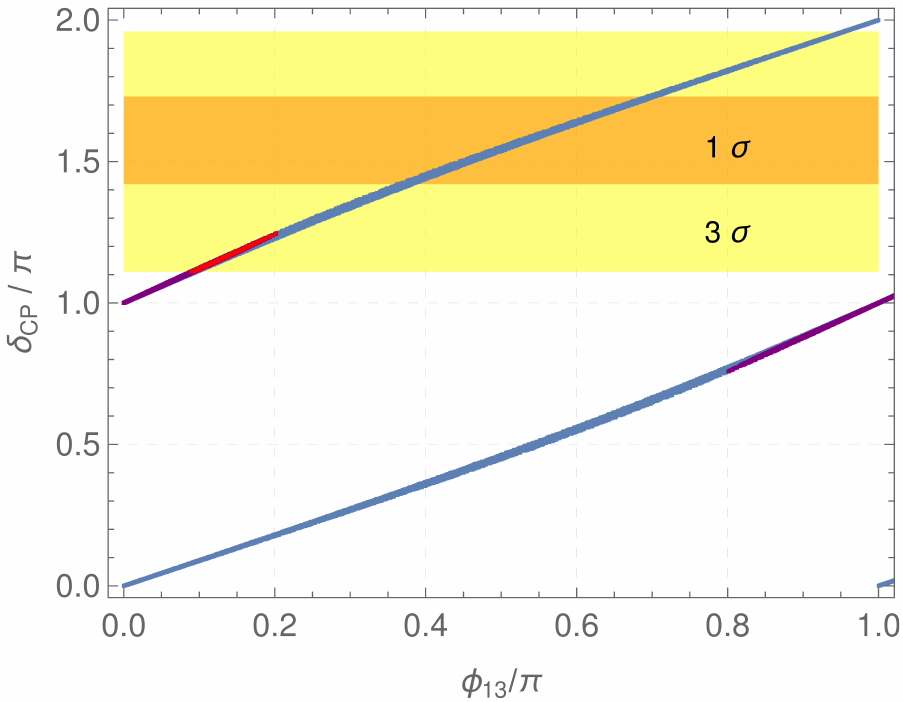} \\
        \includegraphics[width=0.6\textwidth]{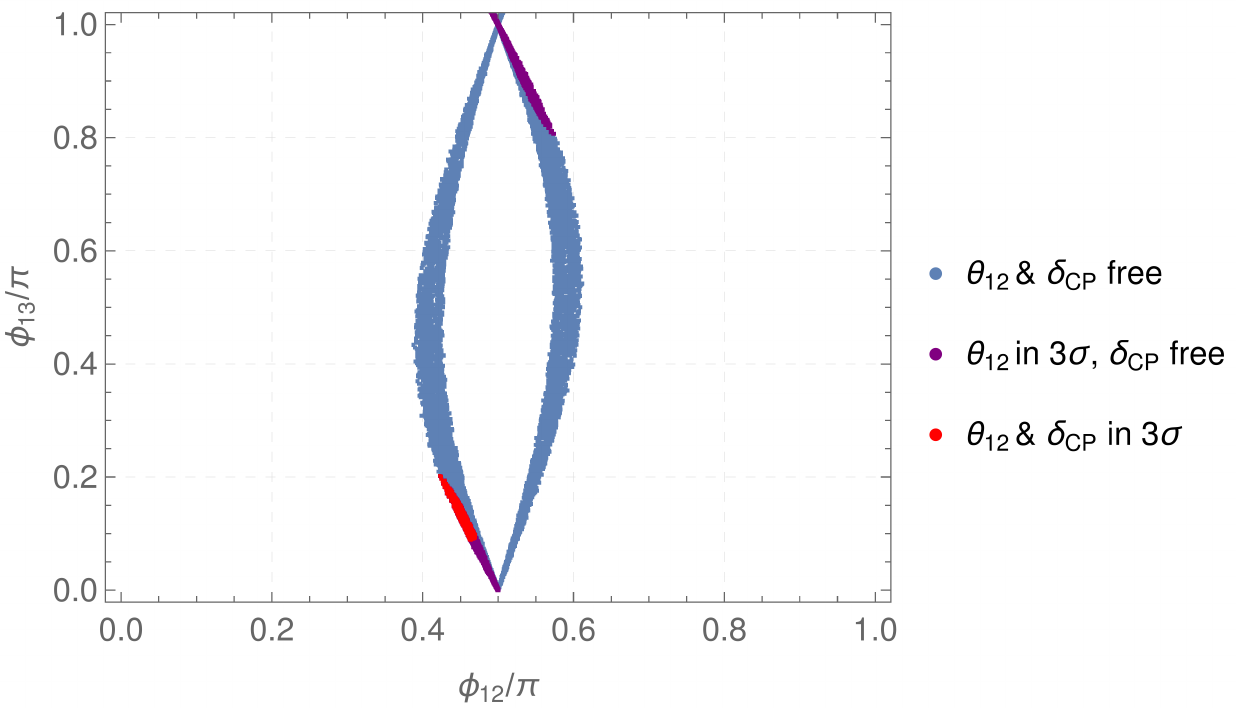}
    \caption{Correlations between physical phases in the IO case. Left up: Correlation between $\phi_{12}$ and $\delta_{CP}$. Right up: Correlation between $\phi_{12}$ and $\delta_{CP}$. Down: Correlation between the Majorana phases $\phi_{13}$ and $\phi_{13}$. In all the plots blue dots arise when the other observables are fitted inside their $3 \sigma$ experimental ranges except for $\theta_{12}$ and $\delta_{CP}$, which are free. In addition, purple dots fit $\theta_{12}$ and red dots also fit $\delta_{CP}$ at the $3\sigma$ level. By imposing all the experimental constraints, the model predicts $\delta_{CP} \approx 1.2 \pi$, $\phi_{12} \approx 0.45 \pi$ and $\phi_{13} \approx 0.12 \pi$ plus a small variance.}
    \label{fig:phases}
\end{figure}

Moreover, we are using the global fit \cite{deSalas:2020pgw} to produce the plots, although the other two global fits \cite{Esteban:2020cvm, Marrone:2015nip} yield slightly lower values for $\delta_{CP}$, thus reducing the tension of the model. Taking $\theta_{12}$ alone, we can see that the model can accommodate $\theta_{12} \approx \theta_{12}^\text{best fit}$ if $\delta_{CP} \approx \pi$. In other words, this model prediction may be tested in the following data releases of neutrino oscillation experiments. Furthermore, the Majorana phases $\phi_{12}$ and $\phi_{13}$, relevant for neutrinoless beta decay experiments, also obtain a strong correlation, as seen in figure~\ref{fig:phases}. The striking similarities between these correlations and the ones in \cite{Chen:2018lsv, Chen:2019fgb} may indicate that our setup leads to the partial conservation of some of the TBM symmetries of the neutrino mass matrix.

\begin{figure}[t]
    \centering
        \includegraphics[width=0.8\textwidth]{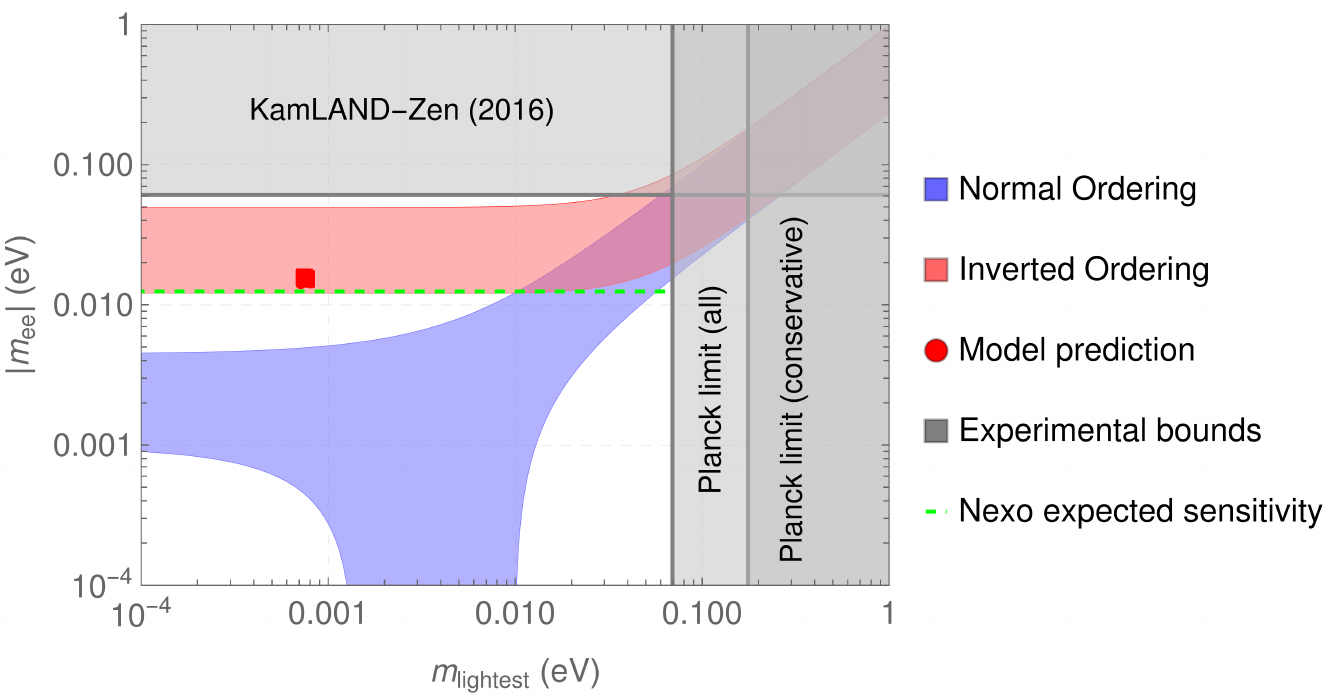}
    \caption{$|m_{ee}|$ is restricted to a small region in this model. The reason is that $m_{light}$ and the Majorana phases are predicted, as well as the ordering. The deviation from a single point comes from the experimental uncertainties in $\Delta m^2_{ij}$ and a small variance in $\phi_{ij}$. The current experimental constraints are given by KamLAND \cite{KamLAND-Zen:2016pfg} and Planck \cite{Planck:2018vyg}. In the future, nEXO is expected to have enough sensitivity to completely rule out the inverted ordering region \cite{nEXO:2021ujk}.}
    \label{fig:mee}
\end{figure}

For neutrinoless double beta decay, if the Majorana neutrino mass mechanism is the dominant contribution to $0\nu\beta\beta$, its rate will be proportional to the quantity $|m_{ee}|$, given by,
\begin{equation}
 \label{eq:mee}
|m_{ee}| = \left|\sum_i U_{ei}^2 m_i\right| = |c_{12}^2 c_{13}^2 m_1 + s_{12}^2 c_{13}^2 e^{2 i \phi_{12}} m_ 2 + s_{13}^2 e^{2 i \phi_{13}} m_3| \, .
\end{equation}
In our model the Majorana phases are approximately fixed as $\phi_{12} \approx 0.45 \pi$ and $\phi_{13} \approx 0.12 \pi$, while the neutrino masses are also predicted to be around $m_3 \approx 7.51 \times 10^{-4}$ eV, $m_1 \approx 4.95 \times 10^{-2}$ eV, $m_2 \approx 5.02 \times 10^{-2}$ eV. Small deviations from these values are possible due to the experimental uncertainty on $\Delta m^2_{ij}$ and the variance in $\phi_{ij}$. This automatically leads to a definite prediction of $|m_{ee}|$ in our model:
\begin{equation}
|m_{ee}^\text{model}| \approx 0.018 \text{ eV} \, ,
\end{equation}
Note that the term with $\phi_{13}$ in \eqref{eq:mee} interferes constructively to $|m_{ee}|$ but is strongly suppressed by $s_{13}^2 m_3$, while the term with $\phi_{12}$ interferes destructively. This is why the allowed points in the model are in the lower region of $|m_{ee}|$ as seen in figure~\ref{fig:mee}. The nEXO experiment is expected to test this model prediction in the future \cite{nEXO:2017nam, nEXO:2021ujk}.

\subsection{Normal ordering}

\begin{figure}[t]
    \centering
        \includegraphics[width=0.7\textwidth]{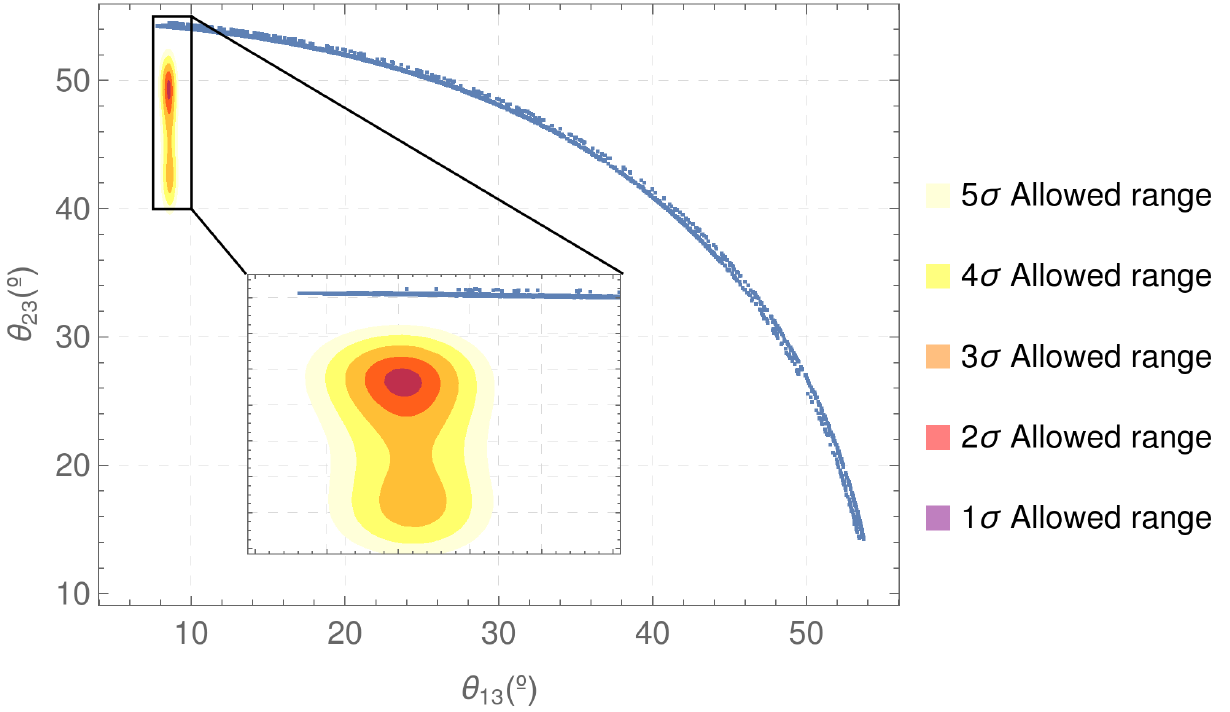}
    \caption{Model prediction in the NO case. The correlation between $\theta_{13}$ and $\theta_{23}$ is incompatible with current experimental constraints at more than $7\sigma$. Therefore the model predicts IO. $\chi^2$ profiles extracted from the global fit \cite{deSalas:2020pgw}.}
    \label{fig:NO13v23}
\end{figure}

In the Normal Ordering case, after imposing the correct charged lepton and neutrino masses at $3\sigma$, a strong correlation appears between the mixing angles $\theta_{23}$ and $\theta_{13}$ in the neutrino sector. This correlation is not compatible with experimental constraints by more than $7\sigma$, as can be seen in figure~\ref{fig:NO13v23}. Therefore, \textbf{Normal Ordering of neutrino masses cannot be realised in this model.}

\section{Dark matter sector} \label{sec:dm}

The $\Sigma(81)$ flavour symmetry has the additional property of stabilizing the lightest of the dark sector fields. In order to see how this mechanism works, we must first note that the singlets $\mathbf{1}_{(i, j)}$ and the $\mathbf{3}_D$, $\mathbf{\bar{3}}_D$ triplets form a closed subset under the tensor products, i.e.
\begin{eqnarray}
    \mathbf{1}_{(i, j)} \times \mathbf{1}_{(k, l)} = \mathbf{1}_{(i+k,\, j+l)},& \hspace{0.5cm}& \mathbf{1}_{(i, j)} \times \mathbf{3(\bar{3})}_D = \mathbf{3(\bar{3})}_D, \\
    \mathbf{3(\bar{3})}_D \times \mathbf{3(\bar{3})}_D = \mathbf{\bar{3}(3)}_D, &\hspace{0.5cm}& \mathbf{3}_D \times \mathbf{\bar{3}}_D = \mathbf{1}_{(i, j)} \, .
\end{eqnarray}

We start by imposing the condition that all of the \textit{visible sector} fields, i.e. the Standard Model fermions and Higgs, transform as either $\mathbf{1}_{(i, j)}$, $\mathbf{3}_D$ or $\mathbf{\bar{3}}_D$. This automatically implies that any effective operator formed by any arbitrary combination of SM particles, $\mathcal{O}_\text{visible}$, will still transform under the same subgroup, i.e. as $\mathbf{1}_{(i, j)}$, $\mathbf{3}_D$ or $\mathbf{\bar{3}}_D$.

Consider now a field $\eta$ belonging to the \textit{dark sector} and transforming as, for example, $\mathbf{3}_A$. It's clear that the effective operator $\eta \, . \, \mathcal{O}_\text{visible}$ cannot be invariant under $\Sigma(81)$, because no operator of the type $\mathcal{O}_\text{visible}$ transforms as $\mathbf{\bar{3}}_A$.

\begin{figure}[t]
    \centering
        \includegraphics[width=0.8\textwidth]{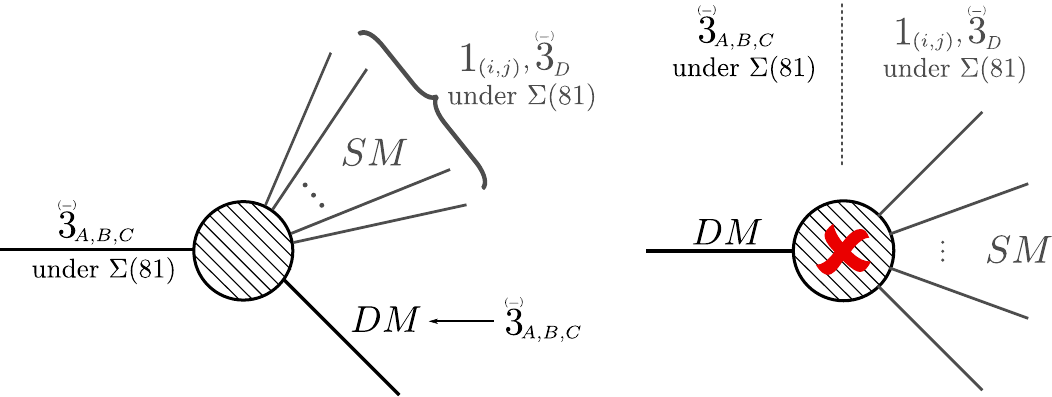}
    \caption{Stability diagrams of the dark matter sector in the model. The lightest particle charged as \textbf{$3(\bar{3})_{A, B, C}$} cannot decay into only Standard Model fields. Left: allowed decay channels for a dark sector particle will necessarily include at least one dark sector particle in the final state (if kinematically allowed). Right: the lightest dark sector particle cannot decay into Standard Model particles due to the flavour symmetry, thus ensuring its stability.} 
    \label{fig:stability}
\end{figure}

In conclusion, any symmetry invariant decay operator of a particle belonging to the dark sector must involve, at least, one dark sector particle in the final state and, thus, the lightest of them will necessarily be stable (see figure~\ref{fig:stability}). In our model the dark matter could be either the lightest neutral mass eigenstate of the scalars $\eta$, $\phi$ or the vector-like fermion $N$, if lighter than the scalars.

Note that this is a generalised, non-Abelian version of the original scotogenic mechanism of \cite{Ma:2006km}, where the stability of the dark matter candidate is enforced by a $\mathbb{Z}_2$ symmetry. This mechanism was extended to Abelian symmetries in \cite{Bonilla:2018ynb,CentellesChulia:2019gic}. Moreover, in \cite{Lavoura:2012cv} the authors present a similar mechanism for some discrete subgroups of $SU(2)$.

While similar to the original scotogenic model, $\Sigma(81)$ makes a clear distinction as already explained, even at the neutrino mass level: there are three independent scotogenic-like diagrams. Each set of \textit{unconnected} fields and couplings were denoted with colours on Eqs.~\eqref{eq:LDY}-\eqref{eq:mnu} for clarity. Regarding the dark matter, this means that the model contains a three-component DM, where the candidates are the lightest of each triad $\{\eta_1, \phi_3, N_2\}$, $\{\eta_2, \phi_1, N_3\}$ and $\{\eta_3, \phi_2, N_1\}$. Of course, after EWSB the neutral part of each scalar doublet will split into its CP-odd and CP-even components due to the usual $\lambda_5$ term in equation \eqref{eq:Vnu}. For more details on the scalar sector, we refer to Appendix \ref{app:scalar}.

It is worth noting that while in most of the Scotogenic scenarios the scalar DM is normally preferred \cite{Avila:2021mwg}, due to the problems of underproduction versus large cLFV for the fermion singlet candidate \cite{Vicente:2014wga}, in this case the issue is largely mitigated. The fact that we have a three-component DM and no large contributions to cLFV, makes the singlet fermion DM case again phenomenologically interesting. Nevertheless, a complete study of the dark matter phenomenology is beyond the scope of this paper.

\section{Conclusions}
\label{sec:conclusions}

We have presented a simple but extremely predictive variant of the scotogenic model. We promoted the scotogenic $\mathbb{Z}_2$ symmetry of the original work to a non-Abelian $\Sigma(81)$ symmetry, which will satisfy the same role of stabilizing the dark matter candidates running in the neutrino mass loop. We considered that leptons, as well as the Higgs doublet $H$, transform as triplets under the flavour symmetry, thus, resembling a 3HDM. These three scalar doublets are responsible for all the spontaneous symmetry breaking, which implies that the model does not need extra flavons in order to fit the experimental data. We found that such a model can, not only satisfy the current experimental constraints, but also lead to very strong and testable predictions in the close future. Fitting the charged lepton masses, $\theta_{13}$ and $\theta_{23}$ inside their $3\sigma$ allowed ranges, we automatically obtained the following predictions:
\begin{itemize}
    \item Neutrino mass sum rule: $\frac{1}{2} \sum m_i = m_\text{heaviest} $.
    
    \item Only Inverted Ordering is realised.
    
    \item These two conditions together lead to $m_\text{lightest} \approx 7.5 \times 10^{-4}$ eV, with some small deviations due to experimental uncertainty in $\Delta m^2_{ij}$.
    
    \item Strong correlation between $\theta_{12}$ and $\delta_{CP}$ as shown in figure~\ref{fig:IO12vdelta}, testable in the near future \cite{Smith:2022ezh,NOvA:2016kwd}.
    
    \item Majorana phases predicted to be around $\phi_{12} \approx  0.45 \pi$ and $\phi_{13} \approx 0.12 \pi$, when all the other observables are in their experimental allowed ranges (see figure~\ref{fig:phases}).
    
    \item The prediction of the Majorana phases and $m_\text{lightest} $ lead to $|m_{ee}| \approx 0.018 \text{ eV}$, testable in future neutrinoless double beta decay experiments \cite{nEXO:2021ujk} (see figure~\ref{fig:mee}).
    
    \item The flavour symmetry $\Sigma(81)$ ensures the stability of the dark matter candidate, which could be either fermionic or scalar. No other symmetries are required apart from the Standard Model gauge symmetries and the spontaneous symmetry breaking comes solely from the three Higgs gauge doublets arranged into a flavour triplet.
\end{itemize}

\begin{appendix}

\section{$\Sigma(81)$ Group}
\label{sec:Sigma81}

The group $\Sigma (81)$ is a discrete, non-Abelian subgroup of $U(3)$ \cite{Jurciukonis:2017mjp} and belongs to the family of groups $\Sigma(3 N^3)$. It has four generators denoted by $a$, $a^{\prime}$, $a^{\prime \prime}$, and $b$, which fulfill the relations,
\begin{equation}
a^3={a^{\prime}}^3={a^{\prime \prime}}^3=1, \qquad a a^{\prime} = a^{\prime} a,\qquad a a^{\prime \prime} = a^{\prime \prime} a,\qquad a^{\prime} a^{\prime \prime} = a^{\prime \prime} a^{\prime}, 
\end{equation}
\begin{equation}
    b^3=1,\qquad b^2 a b = a^{\prime \prime}, \qquad b^2 a^{\prime \prime} b = a^{ \prime}, \qquad b^2 a^{\prime} b = a.
\end{equation}
All the elements of $\Sigma (81)$ can be written in terms of the four generators as,
\begin{equation}
    \forall g \in \Sigma (81), \qquad g = b^{k}a^{n}{a^{\prime}}^{n}{a^{ \prime \prime}}^{l}, \qquad \text{with } \quad k,n,n,l=0,1,2.
\end{equation}
The representations of $\Sigma(81)$ used for the fields multiplets in this model are $\mathbf{3}_{A}$, $\mathbf{\bar{3}}_{A}$, $\mathbf{3}_{D}$, and  $\mathbf{\bar{3}}_{D}$. 
We choose the following basis for these representations:  in the $\mathbf{\bar{3}}_{A}$,
\begin{equation}
b=
\begin{pmatrix} 
0 & 1 & 0 \\
0 & 0 & 1 \\
1 & 0 & 0
\end{pmatrix}, \quad 
a = 
\begin{pmatrix} 
\omega & 0 & 0 \\
0 & 1 & 0 \\
0 & 0 & 1
\end{pmatrix}, \quad 
{a}^{\prime} = 
\begin{pmatrix} 
1 & 0 & 0 \\
0 & 1 & 0 \\
0 & 0 & \omega
\end{pmatrix}, \quad 
{a}^{\prime \prime} = 
\begin{pmatrix} 
1 & 0 & 0 \\
0 & \omega & 0 \\
0 & 0 & 1
\end{pmatrix},
    \label{eq:3ARep}
\end{equation}
where $\omega = e^{i \sfrac{2 \pi}{3} }$. In the $\mathbf{\bar{3}}_{D}$ representation,
\begin{equation}
b=
\begin{pmatrix} 
0 & 1 & 0 \\
0 & 0 & 1 \\
1 & 0 & 0
\end{pmatrix}, \quad 
a = 
\begin{pmatrix} 
\omega^2 & 0 & 0 \\
0 & 1 & 0 \\
0 & 0 & \omega
\end{pmatrix}, \quad 
{a}^{\prime} = 
\begin{pmatrix} 
1 & 0 & 0 \\
0 & \omega & 0 \\
0 & 0 & \omega^2
\end{pmatrix}, \quad 
{a}^{\prime \prime} = 
\begin{pmatrix} 
\omega & 0 & 0 \\
0 & \omega^2 & 0 \\
0 & 0 & 1
\end{pmatrix}.
    \label{eq:3DRep}
\end{equation}
Notice that the generators in the $\mathbf{3}_{A}$ representation are the complex conjugate of the generators in the $\mathbf{\bar{3}}_{A}$, and similarly between the $\mathbf{3}_{D}$, and  $\mathbf{\bar{3}}_{D}$.

It's worth showing explicitly one of the key properties of $\Sigma(81)$ that give rise to dark matter stability in the model presented, i.e. the singlet irreps together with $\bm{3}_D$ and $\bm{\bar{3}}_D$ form a close subgroup. This can be seen by looking at the products \eqref{eq:3d3dbar}-\eqref{eq:13d3dbar2}.
\begin{equation}
   \mathbf{1}_{(k,l)} \times \mathbf{3}_D(\mathbf{\bar{3}}_D) = \mathbf{3}_D(\mathbf{\bar{3}}_D) \, , \quad
   \mathbf{3}_{D} \times \mathbf{3}_{D} = \mathbf{\bar{3}}_{D} + \mathbf{\bar{3}}_{D} + \mathbf{\bar{3}}_{D} \, , \quad
   \mathbf{3}_{D} \times \mathbf{\bar{3}}_{D} = \mathbf{1}_{(k,l)} \, ,
\end{equation}
with $k,l = 0,1,2$.

Expanding in components in the basis defined by \eqref{eq:3ARep} and \eqref{eq:3DRep}, we have the tensor products,
\begin{align}
\begin{pmatrix} x_1 \\ x_2 \\x_3 \end{pmatrix}_{\bm{3}_D} \otimes \begin{pmatrix} y_1 \\ y_2 \\y_3 \end{pmatrix}_{\bm{\bar{3}}_D} = \sum_{k=0,1,2} [(x_1 y_1 + \omega^{2k} x_2 y_2 + \omega^{k} x_3 y_3)_{\mathbf{1}_{(k,0)}} &\oplus (x_2 y_3 + \omega^{2k}x_3 y_1+ \omega^{k} x_1 y_2)_{\mathbf{1}_{(k,2)}} \nonumber\\  &\oplus (x_3 y_2 + \omega^{2k}x_1 y_3+ \omega^{k} x_2 y_1)_{\mathbf{1}_{(k,1)}}]. \label{eq:3d3dbar}
\end{align}
\begin{equation}
\begin{pmatrix} x_1 \\ x_2 \\x_3 \end{pmatrix}_{\bm{3}_D} \otimes \begin{pmatrix} y_1 \\ y_2 \\y_3 \end{pmatrix}_{\bm{3}_D} = \begin{pmatrix} x_1 y_1 \\ x_2 y_2 \\x_3 y_3 \end{pmatrix}_{\bm{\bar{3}}_D} \oplus \begin{pmatrix} x_2 y_3 \\ x_3 y_1 \\ x_1 y_2 \end{pmatrix}_{\bm{\bar{3}}_D} \oplus \begin{pmatrix} x_3 y_2 \\ x_1 y_3 \\ x_2 y_1 \end{pmatrix}_{\bm{\bar{3}}_D},
\label{eq:3d3d}
\end{equation}
\begin{equation}
(x)_{\mathbf{1}_{(k,0)}} \otimes \begin{pmatrix}
y_1 \\ y_2 \\ y_3
\end{pmatrix}_{\bm{3}(\bm{\bar{3}})_D}=
\begin{pmatrix}
x y_1\\ \omega^k x y_2\\ \omega^{2k} x y_3
\end{pmatrix}_{\bm{3}(\bm{\bar{3}})_D}
    \label{eq:13d3dbar0}
\end{equation}
\begin{equation}
(x)_{\mathbf{1}_{(k,1)}} \otimes \begin{pmatrix}
y_1 \\ y_2 \\ y_3
\end{pmatrix}_{\bm{3}(\bm{\bar{3}})_D}=
\begin{pmatrix}
x y_3\\ \omega^k x y_1\\ \omega^{2k} x y_2
\end{pmatrix}_{\bm{3}(\bm{\bar{3}})_D}
    \label{eq:13d3dbar1}
\end{equation}
\begin{equation}
(x)_{\mathbf{1}_{(k,2)}} \otimes \begin{pmatrix}
y_1 \\ y_2 \\ y_3
\end{pmatrix}_{\bm{3}(\bm{\bar{3}})_D}=
\begin{pmatrix}
x y_2\\ \omega^k x y_3\\ \omega^{2k} x y_1
\end{pmatrix}_{\bm{3}(\bm{\bar{3}})_D}
    \label{eq:13d3dbar2}
\end{equation}
The label $\mathbf{1}_{(k,l)}$, with $k,l=0,1,2$, represent the nine different one dimensional irreps of $\Sigma(81)$, being $\mathbf{1}_{(0,0)}$ the invariant singlet. 

For further details of the properties of the $\Sigma(81)$ group and the explicit expressions of the tensor products of dark sector fields we refer the reader to the \textbf{second} edition of the book ``An Introduction to Non-Abelian Discrete Symmetries for Particle Physicists'' \cite{Kobayashi:2022moq}, since the first edition had inconsistencies in the representations used in the tensor products.

\section{Scalar sector} \label{app:scalar}
We now explicitly write down the relevant Lagrangian terms that lead to the masses of the scalars $\eta$ and $\phi$. These are given by equations (\ref{eq:Vnu}) and
\begin{eqnarray} \label{eq:scalarmassetaphi}
    \mathcal{V}_\nu \; &\supset& m_\eta^2 \, (\eta^\dagger \eta) + m_\phi^2 \, (\phi^\dagger \phi) \\
    &+& \sum_{i=1}^3 \lambda_{H\eta i} \left[ (H^\dagger  H) (\eta^\dagger  \eta)\right]_i + \lambda_{H\phi i} \left[ (H^\dagger  H) (\phi^\dagger  \phi)\right]_i \nonumber \\
    &+& \sum_{i=4}^6 \lambda_{H\eta i} \left[ (H^\dagger  \eta) (\eta^\dagger  H)\right]_i + \lambda_{H \phi i} \left[ (H^\dagger  \phi) (\phi^\dagger  H)\right]_i \nonumber
    \, .
\end{eqnarray}
where the sub-index outside a bracket $\left[ X \right]_i$ represent the different contractions under $\Sigma(81)$ to the trivial singlet. Explicitly,
\begin{eqnarray}
m_X^2 (X^\dagger X)& =& m_X^2 (X_1^\dagger X_1+X_2^\dagger X_2+X_3^\dagger X_3) \, , \\
\lambda_{HX1} \left[(H^\dagger H) (X^\dagger X)\right]_1& =& \lambda_{HX1} \left((H_1^\dagger H_1)(X_1^\dagger X_1)+(H_2^\dagger H_2)(X_2^\dagger X_2)+(H_3^\dagger H_3)(X_3^\dagger X_3)\right)  \, , \\
\lambda_{HX2} \left[(H^\dagger H) (X^\dagger X)\right]_2& =& \lambda_{HX2} \left((H_1^\dagger H_1)(X_2^\dagger X_2)+(H_2^\dagger H_2)(X_3^\dagger X_3)+(H_3^\dagger H_3)(X_1^\dagger X_1)\right) \, , \\
\lambda_{HX3} \left[(H^\dagger H) (X^\dagger X)\right]_3& =& \lambda_{HX3} \left((H_1^\dagger H_1)(X_3^\dagger X_3)+(H_2^\dagger H_2)(X_1^\dagger X_1)+(H_3^\dagger H_3)(X_2^\dagger X_2)\right) \, , \\
\lambda_{HX4} \left[(H^\dagger X) (X^\dagger H)\right]_4& =& \lambda_{HX4} \left((H_1^\dagger X_1)(X_1^\dagger H_1)+(H_2^\dagger X_2)(X_2^\dagger H_2)+(H_3^\dagger X_3)(X_3^\dagger H_3)\right) \, , \\
\lambda_{HX5} \left[(H^\dagger X) (X^\dagger H)\right]_5& =& \lambda_{HX5} \left((H_1^\dagger X_2)(X_2^\dagger H_1)+(H_2^\dagger X_3)(X_3^\dagger H_2)+(H_3^\dagger X_1)(X_1^\dagger H_3)\right) \, , \\
\lambda_{HX6} \left[(H^\dagger X) (X^\dagger H)\right]_6& =& \lambda_{HX6} \left((H_1^\dagger X_3)(X_3^\dagger H_1)+(H_2^\dagger X_1)(X_1^\dagger H_2)+(H_3^\dagger X_2)(X_2^\dagger H_3)\right) \, , 
\end{eqnarray}
with $X \in \{\eta, \phi\}$.

As usual, the scalars $\eta_i$ and $\phi_i$ can be written down into their $SU(2)_L$ components as,
\begin{eqnarray}
\eta_i &=& \left(\eta_i^+, \,\, \frac{1}{\sqrt{2}}( \eta_{Ri} + i\, \eta_{Ii})\right)^T \ ,\\
\phi_i &=& \left(\phi_i^+, \,\, \frac{1}{\sqrt{2}}( \phi_{Ri} + i\, \phi_{Ii})\right)^T  \, .
\end{eqnarray}

After EWSB, the charged components of each $\eta_i$ and $\phi_i$ remain as mass eigenstates, with masses
\begin{eqnarray}
m_{\eta_i^+} &\approx& m_\eta^2 + v_1^2 \lambda_{H\eta i},
\\
m_{\phi_i^+} &\approx& m_\phi^2 + v_1^2 \lambda_{H\phi i},
\end{eqnarray}
in the limit $v_1 \gg v_2, v_3$ (see \eqref{eq:vev align}).

On the other hand, the neutral components mix through the $\lambda_5^{(k)}$ terms given in equation \eqref{eq:Vnu}. Assuming a CP conserving scalar potential, which implies $\lambda_5^{(1)}, \lambda_5^{(2)} \in \text{Reals}$, the mixing only happens among CP-odd and CP-even scalars and in pairs $\{\eta_1, \phi_3\}$, $\{\eta_2, \phi_1\}$ and $\{\eta_3, \phi_2\}$. Respectively, the mass matrices for each pair is given by
\begin{eqnarray}
m_{R,I1} \approx \left(\begin{matrix} m_\eta^2 + v_1^2 (\lambda_{H\eta1}+\lambda_{H\eta4})   & & \pm v_1 v_2 \lambda_5^{(2)} \\
  \pm v_1 v_2 \lambda_5^{(2)} & & m_\phi^2 + v_1^2 (\lambda_{H\phi3}+\lambda_{H\phi6})
\end{matrix}\right) \, ,\\
m_{R,I2} \approx \left(\begin{matrix} m_\eta^2 + v_1^2 (\lambda_{H\eta2}+\lambda_{H\eta5})   & & \pm v_1^2 \lambda_5^{(1)} \\
 \pm v_1^2 \lambda_5^{(1)} && m_\phi^2 + v_1^2 (\lambda_{H\phi1}+\lambda_{H\phi4})
\end{matrix}\right) \, , \\
m_{R,I3} \approx \left(\begin{matrix} m_\eta^2 + v_1^2 (\lambda_{H\eta3}+\lambda_{H\eta6})   & & \pm v_1 v_3 \lambda_5^{(2)} \\
  \pm v_1 v_3 \lambda_5^{(2)} & & m_\phi^2 + v_1^2 (\lambda_{H\phi2}+\lambda_{H\phi5})
\end{matrix}\right)  \, .
\end{eqnarray}

By choosing appropriate signs for the $\lambda$ couplings one can make sure that a neutral component is the lightest field of each set.
\end{appendix}

\section{Acknowledgements}

The authors want to thank Andreas Trautner for double covering us with wisdom and Rahul Srivastava for helpful comments. We also thank professor Luís Lavoura and professor Martin K. Hirsch for helpful insight. O.M. is supported by the Spanish grant PID2020-113775GB-I00 (AEI/10.13039/501100011033) and Programa Santiago Grisol\'{i}a (No. GRISOLIA/2020/025). R.C. is supported by the Alexander von Humboldt Foundation Fellowship.

\bibliography{Biby/biby.bib, Biby/bibliography1.bib, Biby/bibliography2.bib, Biby/bibliography3.bib, Biby/bibliography4.bib, Biby/bibliography5.bib, Biby/bibliography6.bib, Biby/mine.bib, Biby/bib-deberes.bib}
\bibliographystyle{utphys}

\end{document}